\documentclass[a4paper,11pt]{article}
\pdfoutput=1
\usepackage{jinstpub} 
\usepackage{graphicx}
\usepackage{subcaption}
\usepackage{float}
\usepackage[linesnumbered,ruled,vlined]{algorithm2e}

\usepackage{lineno}
\usepackage{soul}
\usepackage{comment}

\title{CONCERTO: Characterization of analog readout electronics}

\author[a,1]{M. Abdkrimi,\note{Corresponding author}}
\author[b]{O. Rossetto,}
\author[b]{O. Bourrion,}
\author[b]{C. Hoarau,}
\author[b]{C. Vescovi}

\affiliation[a] {Univ. Grenoble Alpes, CNRS, Grenoble INP, Institut Néel, Grenoble, 38000, France }

\affiliation[b] {Univ. Grenoble Alpes, CNRS, Grenoble INP, LPSC-IN2P3, Grenoble, 38000, France }

\emailAdd{mounir.abdkrimi@gmail.com}

\abstract{
CONCERTO is a millimeter-wave imaging instrument that operated on the Atacama Pathfinder Experiment (APEX) telescope from April 2021~to May 2023.
Its primary scientific objectives include the study of galaxy clusters through the Sunyaev–Zel’dovich (SZ) effects, the observation of Galactic star-forming regions, and the first measurements constraining the power spectrum of dusty star-forming galaxies.
The instrument consists of two detector arrays, each comprising 2400~Microwave Kinetic Inductance Detectors (MKIDs). 
Each of the two arrays comprises six feed-lines and is read out by six KID\_READOUT electronic boards, each capable of reading out one feed-line coupled to 400~frequency-multiplexed MKIDs.
As the demand for higher-resolution millimeter-wave imaging continues to grow, future instruments aim to significantly increase the pixel count, with more than 800~detectors per feed-line.
However, the MKID readout electronics chain is inherently complex, making it difficult to fully understand its performance limits and optimization margins.
To address this challenge, we initiated a modeling effort that first focused on the digital section of KID\_READOUT.
In this phase, we developed a digital twin of the FPGA-based signal processing chain, which led to substantial performance improvements. 
The present paper extends this modeling strategy to the analog readout chain.
It presents the characterization and behavioral modeling of all analog components, allowing us to identify the elements that limit the frequency multiplexing factor, determine the dominant noise contributors, and highlight areas for improvement in signal conditioning for the future-generation board.
Together, these developments establish, to the best of our knowledge, the first consolidated digital-and-analog behavioral framework of an MKID readout architecture, implemented in a unified Python-based modeling environment.

}


\begin{document}
\maketitle
\flushbottom

\section{Introduction}

\begin{figure}[h]
\centering
\begin{subfigure}[b]{0.6\textwidth}
    \centering
    \includegraphics[angle=0,width=\textwidth]{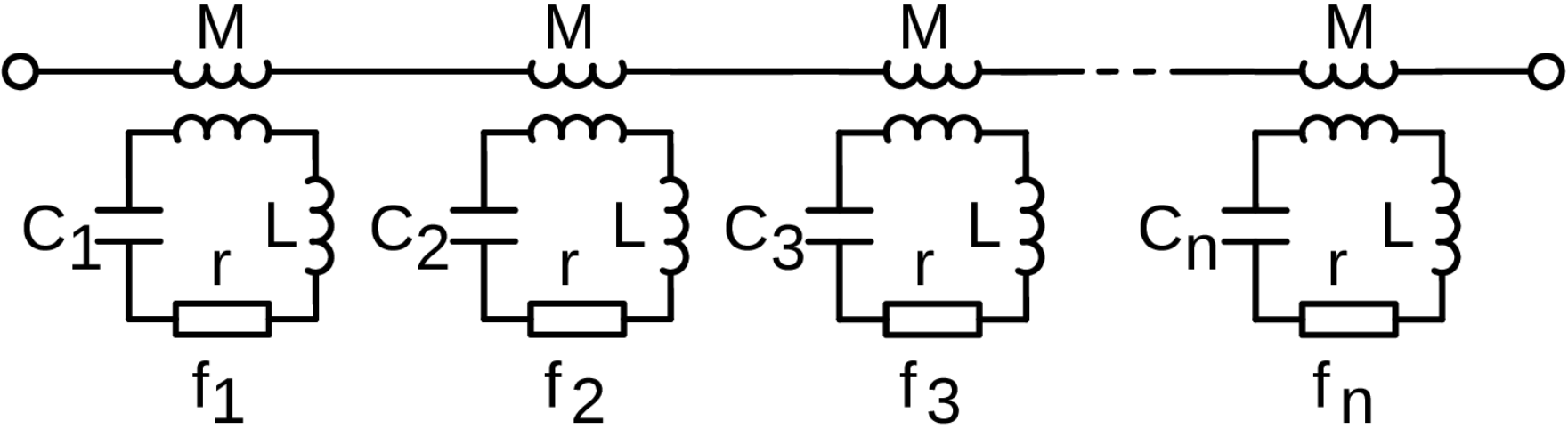}
    \caption{} 
\end{subfigure}
\hfill
\begin{subfigure}[b]{0.39\textwidth}
    \centering
    \includegraphics[angle=0,width=\textwidth]{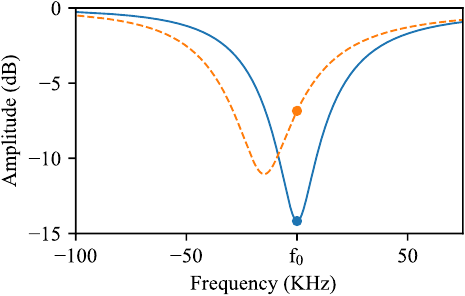}
    \caption{} 
\end{subfigure}
\caption{\label{fig:modelkid} (a) The electrical model of multiple detectors connected in series, with each resonator having a unique resonant frequency ($f_1$, $f_2$, $f_3$, ..., $f_n$). (b) The frequency response of one MKID resonator before excitation (solid blue) and after excitation (dashed orange).}
\end{figure}

Microwave Kinetic Inductance Detectors (MKIDs) have experienced notable advancements over the past twenty years, emerging as a promising detector for observing millimeter-wavelength photons in astronomical research~\cite{baselmans2012kinetic, doyle2008lumped, mazin2005microwave}.
As illustrated in Fig.~\ref{fig:modelkid}, these detectors are LC resonators, with their inductance sensitive to mm-waves, operating in the radiofrequency (RF) domain at cryogenic temperatures around 100\,mK. 
Thanks to the high quality factor of MKIDs, it is possible to connect a large number of MKID resonators in series, each with a unique resonance frequency.

\begin{figure}[h]
\centering
\begin{subfigure}[b]{0.4\textwidth}
    \centering
    \includegraphics[angle=0,width=\textwidth]{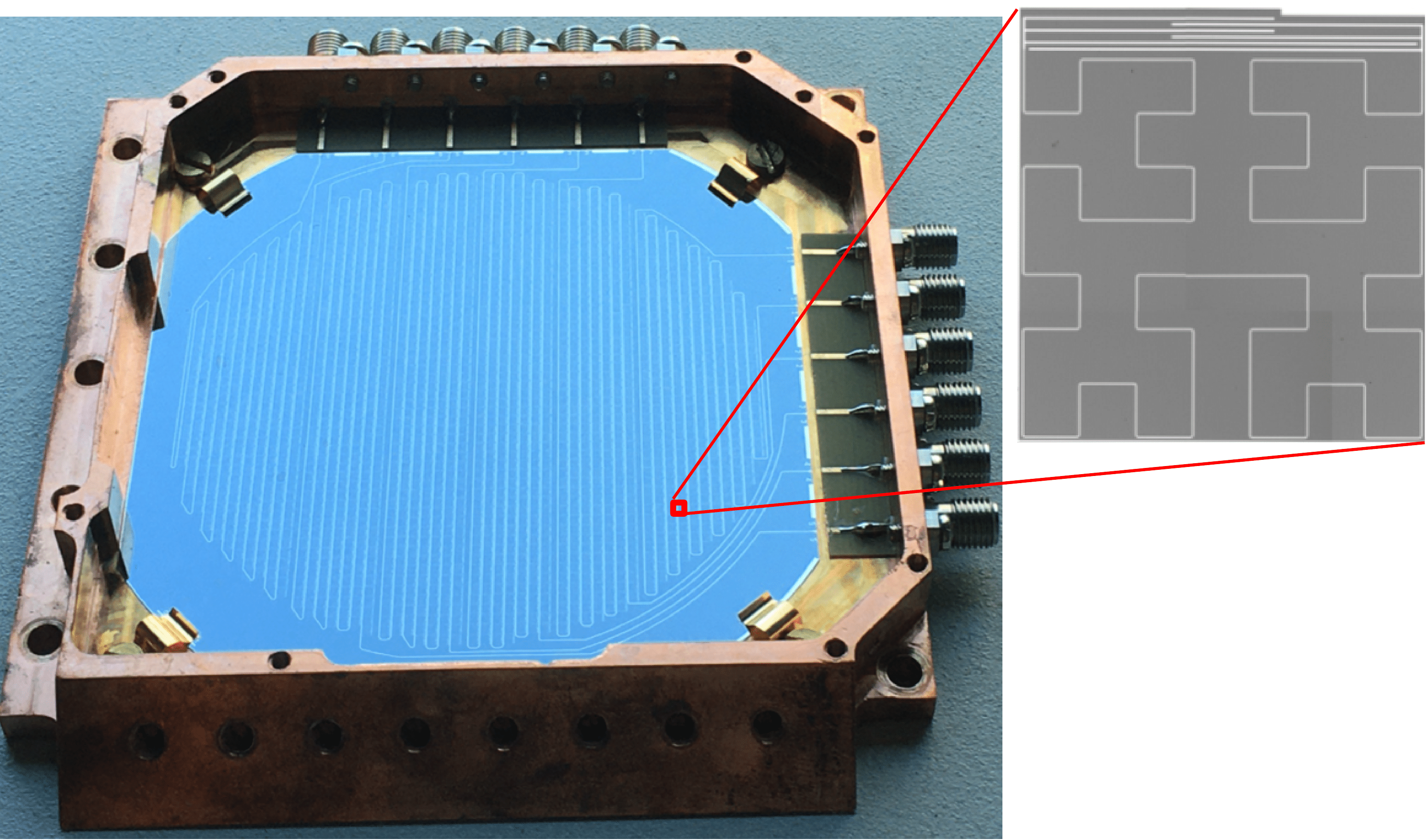}
    \caption{} 
\end{subfigure}
\hfill
\begin{subfigure}[b]{0.59\textwidth}
    \centering
    \includegraphics[angle=0,width=\textwidth]{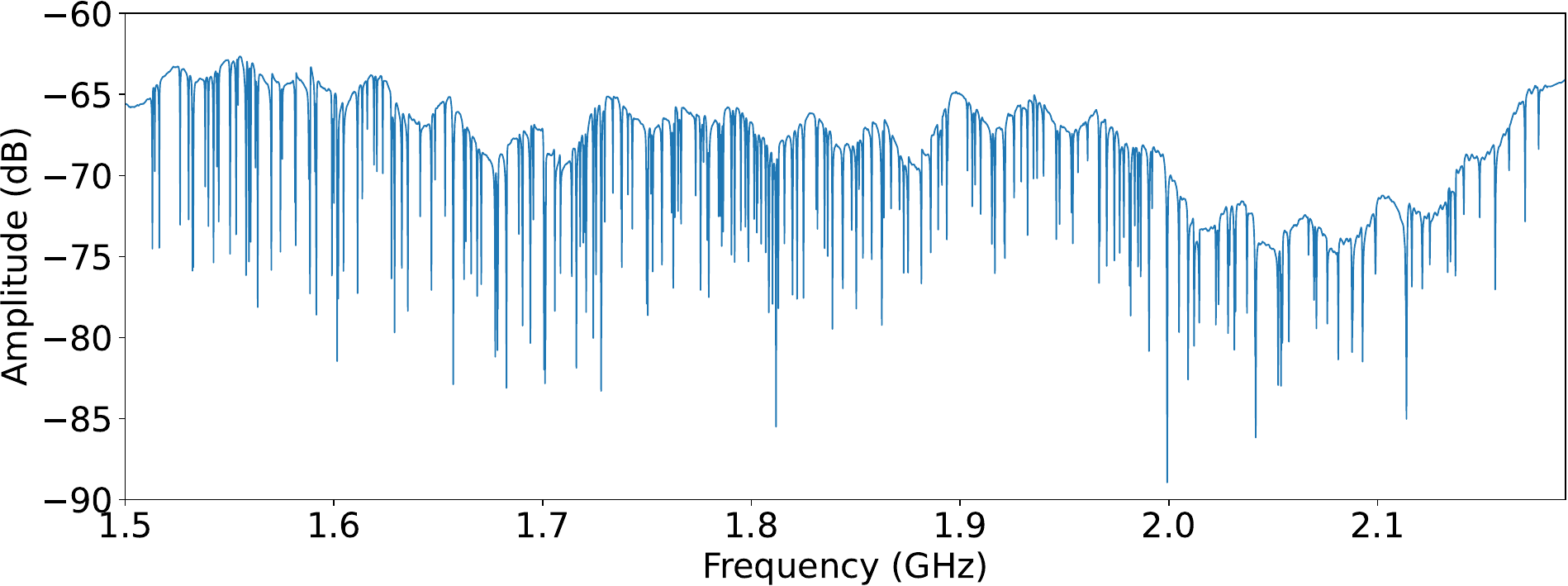}
    \caption{} 
\end{subfigure}
\caption{\label{fig:s21etkid} (a) Photograph of Concerto's MKID array.
(b) Frequency response in transmission of one feed line.}
\end{figure}

Fig.~\ref{fig:s21etkid} shows MKIDs array arranged in a matrix consisting of 6~feed-lines, forming an imager in mm-wave domain, with 2400~RF resonators.
It also illustrates the frequency response of one of CONCERTO's feed-line coupled to 400~resonators having their self-resonant frequencies distributed across a bandwidth of 1\,GHz, with an average separation of 2\,MHz~\cite{concerto1,concerto2}. 

The developed readout electronics board, KID\_READOUT, combines advanced digital frequency multiplexing with analog RF and low-noise conditioning techniques and is currently implemented in the CONCERTO
instrument~\cite{bourrion2022concerto}.

Despite the increasing complexity of MKID readout systems, the state of the art in dedicated simulation tools remains relatively limited. A previous study has focused on partial aspects of the readout chain, such as the analog section alone, using RF Blockset in MATLAB~\cite{kim2025modeling}. Comprehensive modeling approaches that integrate both digital and analog stages are still scarce.

This motivated the development of a unified simulation tool for the KID\_READOUT electronics. The effort began with the digital section, where we developed a Python-based digital twin of the complete digital processing chain currently implemented on KID\_READOUT. This model is both bit-accurate and cycle-accurate with respect to the hardware implementation. It enabled significant reductions in digital resource consumption without degrading readout performance. These results have been presented in previous works~\cite{abdkrimi2025efficient,abdkrimi2025cordic,abdkrimi2026spurs}.

We then extended the modeling effort to the analog part of the system.
An initial study focused on the DAC used in KID\_READOUT, which provided insight into its limitations in tone multiplexing~\cite{abdkrimi2024modeling}. In the present paper, we further expand this work to the full analog readout chain.

To the best of our knowledge, the work presented in this paper represents the first step toward a consolidated behavioral framework of the complete MKID readout chain, developed in Python within a unified modeling environment.
Although the model is based on the current KID\_READOUT architecture, all its constituent blocks—both digital and analog—are fully parameterizable and can be adapted for the characterization and performance evaluation of next-generation readout systems, including alternative digital algorithms and different choices of analog components.



\section{Overview of KID\_READOUT and readout methodology}
\label{subsec:hardaware}
The KID\_READOUT board consists of two main parts: the back-end and the front-end electronics.  
The back-end includes a Field Programmable Gate Array (FPGA) (Xilinx XCKU060FFVA1156-2), a 12-bit Analog-to-Digital Converter (ADC) (ADC12D1000), and a dual 16-bit Digital-to-Analog Converter (DAC) (AD9136).  
Both converters operate at 2 GSamples/s. 
The back-end also features two low-pass filters (LPFs).  
The front-end comprises a quadrature modulator (ADL5375), a mixer (AD8342), and two additional LPFs.  
KID\_READOUT is described in greater detail in a previous work~\cite{bourrion2022concerto}.

\begin{figure}[h]
\centering
\includegraphics[angle=0,width=1\textwidth]{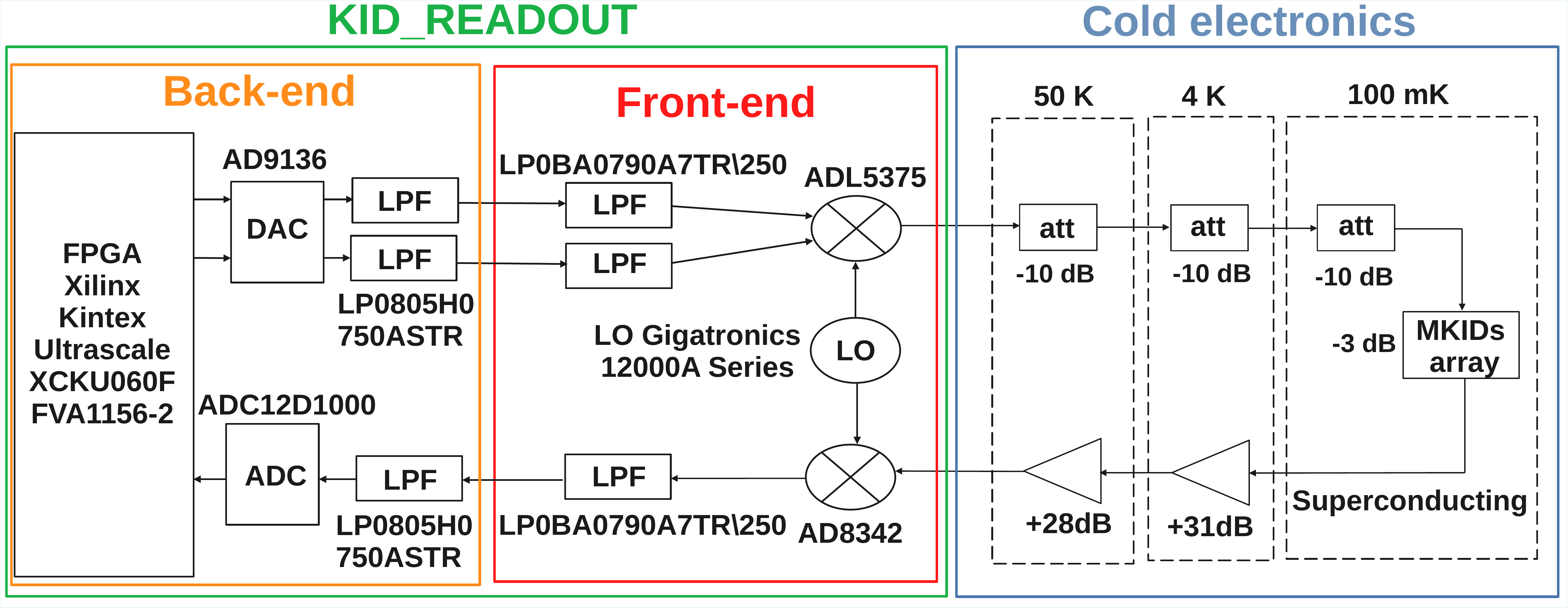}
\caption{\label{wholesys} Overview of the complete electronics setup chain, including components of back-end, front-end, and cold electronics.
}
\end{figure}

The readout methodology involves numerically generating a baseband excitation signal on the FPGA using a Coordinate Rotation Digital Computer (CORDIC).
This signal features a frequency comb covering 1\,GHz bandwidth, where each sinusoid, referred to as a tone, is precisely tuned to probe the resonance frequency (\(f_0\)) of a specific resonator in the array under resting conditions.
This signal is then converted to an analog signal by the DAC and filtered by LPFs (See Fig.~\ref{wholesys}). 
The analog signal is then modulated from the baseband to the MKIDs' resonance band, which lies between 1.2 and 2.2\,GHz, using the quadrature modulator and the frequency generator Gigatronics 12000A Series, which generates a 1.2\,GHz local oscillator (LO).
To minimize added noise and adjust the signal power according to the resonators' requirements, the signal passes through attenuators at various temperature stages as illustrated in Fig.~\ref{wholesys}.

The signal exiting the MKIDs array is amplified by low noise amplifiers, and then
forwarded to the warm electronics, where it is demodulated back to the baseband range using the mixer and the same LO used during modulation. 
After low-pass filtering, the signal is fed into ADC. 
The digitized signal covers the same baseband as the excitation signal, specifically [0-1\,GHz].
The returning frequency comb is then down-converted and analyzed by channelized Digital Down Converters (DDCs), which extract In-phase (I) and Quadrature (Q) components of each tone present in the frequency comb. 
The I and Q components are utilized to compute the amplitude, given by \( \sqrt{I^2 + Q^2} \), and the phase, given by \( \arctan\left(\frac{Q}{I}\right) \), permitting the deduction of the energy absorbed by each pixel within the detector.

\section{Analog chain characterization and modeling}

In this study, four components are characterized and modeled: two on the back-end card (ADC, DAC) and two on the front-end board (IQ modulator, mixer).
It is essential to note that all components used are off-the-shelf market products, whose internal architectures are not disclosed. 
Thus, we developed empirical behavioral models that replicate the observed imperfections.
The spectra produced by both measurements and models are then compared for model validation.

This modeling effort serves multiple complementary purposes. 
First, while the current KID\_READOUT multiplexes 400~detectors per feedline, it has never been determined whether this represents the true maximum multiplexing factor. 
This modeling explicitly identifies which component—if any—sets this limit and quantifies the system's true multiplexing capability.
Second, it identifies the dominant noise source among the analog components.
Finally, it characterizes the spurs generated by these components, which may propagate through the electronics chain and corrupt the I/Q signals of one or more of the 400~readout channels, thereby compromising the accuracy of the extracted amplitude, phase, and energy absorption measurements.

For the measurement setup, the KID\_READOUT comprises two interconnected cards: the back-end electronics card and the RF mezzanine card. 
The RF card comes in two variants: the basic version, which is used to read out the MKID arrays and includes the modulator and demodulator, and the debug version, which uses only baluns and does not up/downconvert the signal.

\subsection{Digital-to-Analog Converter}

The AD9136 is a 16-bit DAC operating at a sampling frequency of 2\,GHz. 
As a commercial off-the-shelf component, it was not specifically designed for our application, which requires generating a superposed 400-tone signal that spans a large 1\,GHz bandwidth.

In this context, we investigated the impact of multi-tone superposition on the maximum achievable multiplexing factor as a function of the DAC’s full-scale range and slew-rate limitation.

In addition, we modeled its Differential Non-Linearity (DNL), Integral Non-Linearity (INL), and white noise in order to reproduce the measured noise floor and harmonic distortion.
This allowed us to assess its contribution to the overall noise of the complete readout chain and determine whether it constitutes a dominant noise source.

\subsubsection{INL, DNL and white noise} \label{subsec:dac_inl_dnl}

The DNL specification of the AD9136 DAC is $\pm 1$~least significant bit (LSB).  
This means that for each digital input code applied to the DAC, the resulting output voltage may deviate by up to one LSB from the ideal step size.  
Consequently, the relationship between the input code and the output voltage deviates from the ideal linear form, given by $\text{code} \times V_s$, where $V_s$ is the voltage step size defined as $1\,\mathrm{V}_{\mathrm{pp}}/2^{16}$, with $1\,\mathrm{V}_{\mathrm{pp}}$ representing the full-scale output range of the AD9136. Accordingly, the output voltage is modeled as:

\begin{equation}
\mathrm{Voltage}_{\mathrm{out}}[\mathrm{code}]
=
\mathrm{code}\times V_s
+
\mathrm{DNL}_{\mathrm{error}}[\mathrm{code}]
\label{eq:voltage-out-dnl}
\end{equation}

where $\text{DNL}_{\text{error}}$ is bounded within the range $[-1\,\text{LSB} \times V_s,\,+1\,\text{LSB} \times V_s]$, such that the cumulative sum of DNL errors, which defines the INL, remains within the AD9136's INL specification of $\pm 1$~LSB.
  
The developed model was simulated using a sinusoidal input signal, with a representative frequency of 150\,MHz and a full-scale amplitude of $2^{15}$, consistent with the DAC’s dynamic range of \([-2^{15},\, 2^{15}-1]\). 
The simulation was performed using a sampling frequency of 2000\,MHz—matching that of the DAC—and a frequency resolution of 8\,kHz.

For an ideal $N$-bit converter driven by a full-scale sine wave, the quantization noise floor over a frequency bin of resolution $\Delta f$ is given by:

\begin{equation}
\mathrm{Noise\ Floor}_{\mathrm{dBc}}
=
-6.02N - 1.76
-10\log_{10}\!\left(\frac{f_s}{2}\right)
+10\log_{10}(\Delta f)
\label{eq:quantization-noise-floor}
\end{equation}

where white quantization noise is assumed to be uniformly distributed over the Nyquist bandwidth $[0, f_s/2]$.
For $N = 16$, $f_s = 2000$ MHz, and $\Delta f = 8$ kHz, this yields a theoretical noise floor of $-149.05$\,dBc.
This value represents the DAC's noise floor considering only quantization noise.

\begin{figure}[h]
    \centering
    \includegraphics[width=0.65\textwidth]{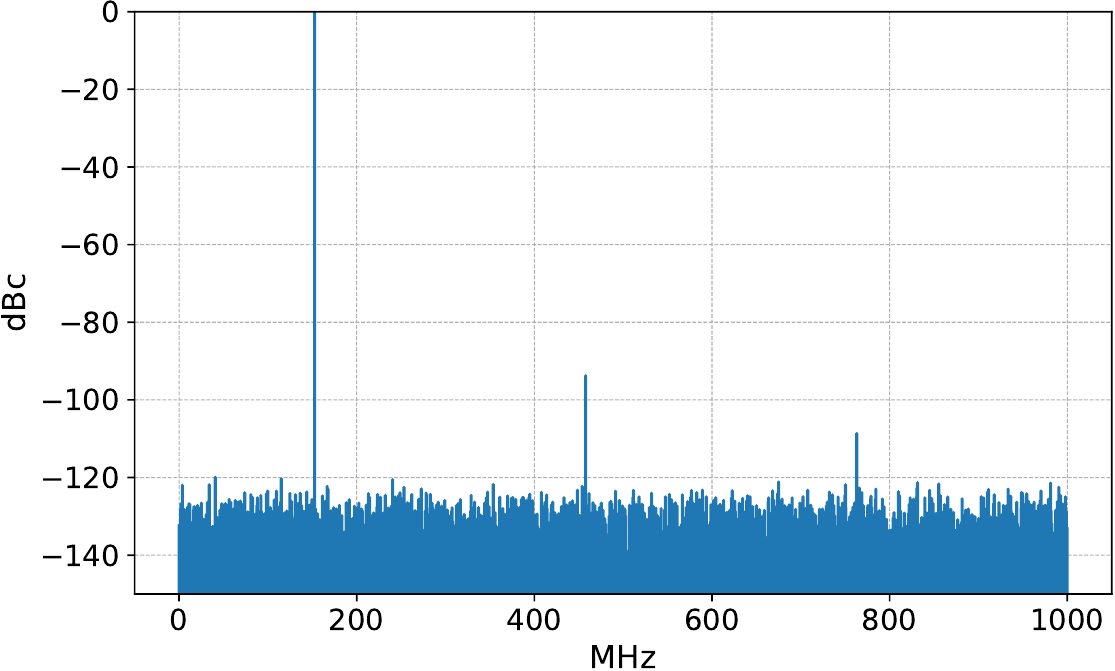}
    \caption{
Output spectrum from the simulation of the developed DAC model incorporating DNL and INL imperfections.}
    \label{fig:DAC_INL}
\end{figure}

However, the model simulation results in a spectrum with an elevated noise floor, reaching approximately $-124\,\text{dBc}$, which is about 25\,dB above the theoretical quantization limit of $-149.05\,\text{dBc}$, as shown in Fig.~\ref{fig:DAC_INL}.

Although this level could not be verified experimentally due to the spectrum analyzer's sensitivity limit of approximately $-100$\,dBm, the simulated noise is consistent with the datasheet, which specifies a spectral noise floor of $-163$\,dBc/Hz. 
By integrating over the 8\,kHz simulation resolution:
\begin{equation}
\mathrm{Noise\ Floor}_{\mathrm{dBc}}
=
-163~\mathrm{dBc/Hz}
+
10\log_{10}(8000)
=
-163 + 39.03
=
-123.97~\mathrm{dBc}
\label{eq:measured-noise-floor}
\end{equation}

This integrated value aligns closely with the observed simulation noise floor of approximately $-124$\,dBc. 

This confirms that the model effectively degrades the noise performance of the ideal 16-bit DAC and accurately reproduces the noise floor specified in the datasheet.

Beyond white noise, the model also reproduces odd-order harmonic distortion resulting from a realistic modeling of DAC nonlinearity through INL and DNL.
Specifically, the third harmonic appears at 93\,dB below the fundamental tone, and the fifth harmonic at 108\,dB below.
  
However, in our MKIDs readout scenario, these components are not observable, as the tones have power levels on the order of –32\,dBm each, placing the corresponding harmonics below the noise floor.

We therefore conclude that the harmonic impact is negligible. 
Additionally, the contribution to the noise floor is also negligible, as it is masked by the noise floor of the 12-bit ADC (see Section~\ref{subsec:adc_inl_dnl}).

\subsubsection{DAC clipping and Slew rate}

This section presents results obtained by modeling both the DAC slew rate and its finite digital dynamic range;
detailed analyses are provided in~\cite{abdkrimi2025modeling}. 
The objective is to determine the maximum multiplexing factor, defined as the maximum number of tones that can be generated without exceeding either the DAC digital input range or the slew-rate limit.

\begin{figure}[h]
    \centering
    \includegraphics[width=0.7\textwidth]{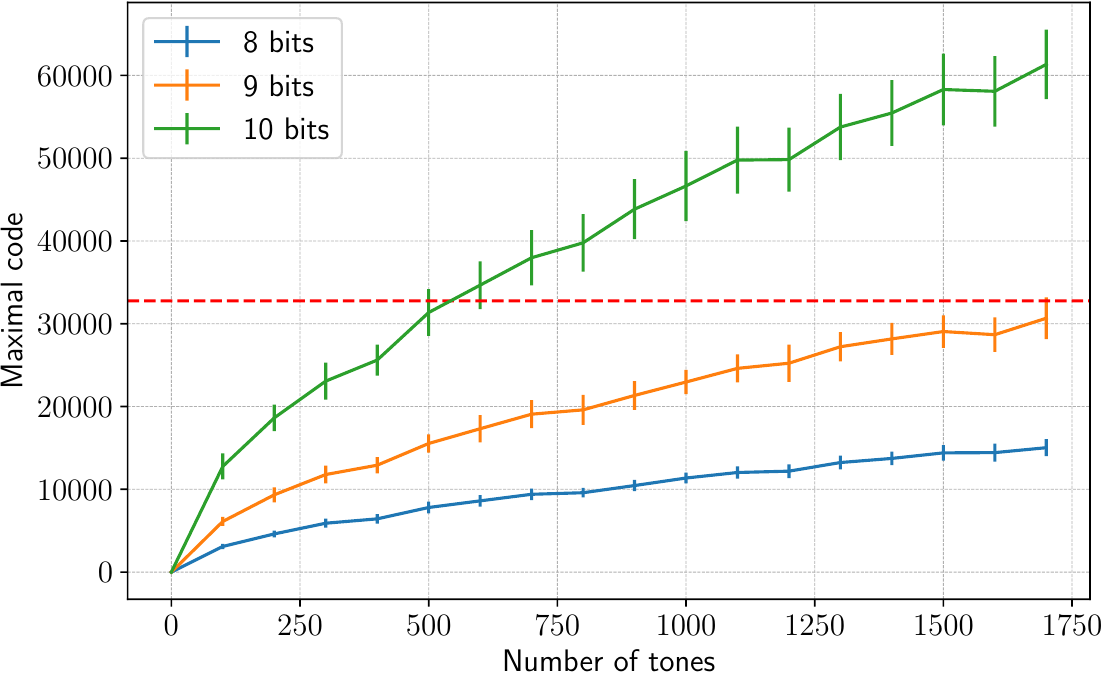}
    \caption{Maximum absolute code of the simulated multi-tone signal as a function
    of the multiplexing factor (0--1750 tones) and tone amplitude bit width (8--10 bits).
    The red dashed line marks the DAC's maximum representable code $|2^{15}|$.
    Each point is averaged over 50 simulation runs with random initial phases;
    error bars indicate the corresponding statistical spread.
    Detailed analyses are provided in~\cite{abdkrimi2025modeling}.}
    \label{fig:satur_max}
\end{figure}

As shown in Fig.~\ref{fig:satur_max}, with the current KID\_READOUT configuration
of 10~bits per tone, the multiplexing factor is limited to 400~tones by the DAC
digital input range ($|2^{15}|$). Reducing the number of bits per tone relaxes this
constraint: using 9~bits per tone, for instance, allows up to 1{,}700~tones before
saturation occurs.

\subsection{IQ Modulator}
\subsubsection{Principle and imperfections}

An analog IQ modulator modulates a baseband signal \( S \) with a high-frequency sinusoidal signal known as the local oscillator (LO).
To achieve IQ modulation, both of these signals must be in-phase (I) and quadrature (Q) configuration, which is characterized by a 90-degree phase difference between them.
Upon multiplication and summation, the IQ modulator produces a modulated carrier signal. 
A key characteristic of the IQ modulator is its ability to generate only the desired upper sideband image, unlike conventional modulators that produce both upper and lower sidebands, and require additional filtering to suppress the unwanted lower sideband image.

The specific modulator under characterization, the ADL5375, requires the I and Q components of the signal $S$, and a single-ended LO input, which is then internally converted into differential I/Q LO signals.
This transformation process introduces amplitude imbalances between the LO's I and Q components. 
Moreover, the analog multiplier employed for modulation exhibits inherent non-linearities, typical of such components, which result in the generation of intermodulation products (IMDs)~\cite{peng1995nonlinear,chehrazi2009second}.

All these imperfections manifest as various spurious components in the output spectrum of the IQ modulator, including harmonics of the input signals (S and LO), IMDs (linear combinations of input signals frequencies), feed-through images of the input signals, and non-rejected low sideband image.

In conventional communication systems, transmitted signals typically occupy a narrow bandwidth. 
Modulator spurs are usually suppressed using bandpass filters, which isolate the desired modulated band and remove unwanted spectral components. 
However, in our case, a deliberate decision has been made not to include filtering at the modulator's output. 
Consequently, the excitation signal contains all unwanted harmonics and spurious spectral components generated by the modulation process.
This choice allows flexibility in adjusting the LO frequency, either higher or lower, relative to the frequency range of the instrumented MKIDs feedline.

\subsubsection{Experimental setup and measurements}
\label{subsec:experimental_setup_iq}

\begin{figure}[h]
    \centering
    \includegraphics[width=1\textwidth]{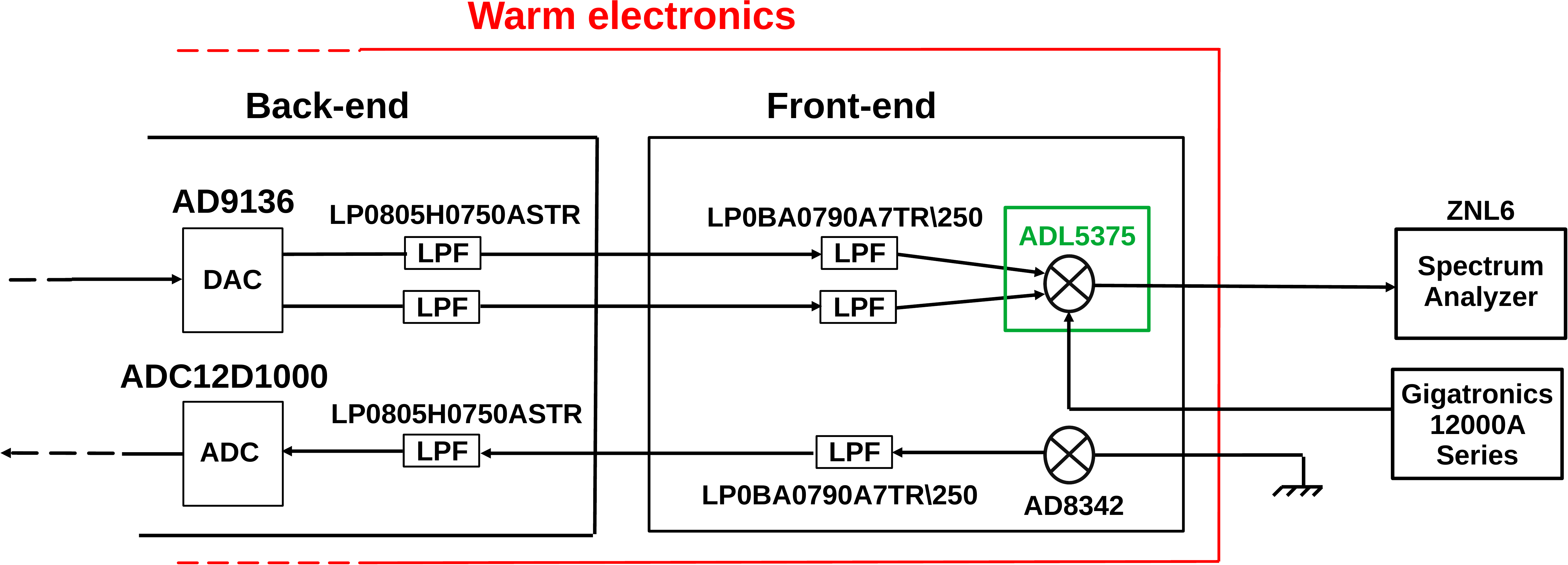}
    \caption{ADL5375 IQ modulator Measurement setup.}
    \label{fig:IQ_setup}
\end{figure}

The ADL5375 quadrature modulator used in the mezzanine card takes as its inputs the signals \(\text{S}_{\text{I}}\) and \(\text{S}_{\text{Q}}\) generated by the KID\_READOUT's FPGA and converted to analog by the DAC, and the \(\text{LO}\) signal generated by the frequency generator Gigatronics 12000A Series.
As mentioned previously, this modulator internally generates the quadrature signals \(\text{LO}_{\text{I}}\) and \(\text{LO}_{\text{Q}}\) from the single input signal \(\text{LO}\). 
Fig.~\ref{fig:IQ_setup} shows the ADL5375 measurement setup, using the mezzanine card version that includes the IQ modulator.

\begin{figure}[h]
    \centering
    \includegraphics[width=0.85\textwidth]{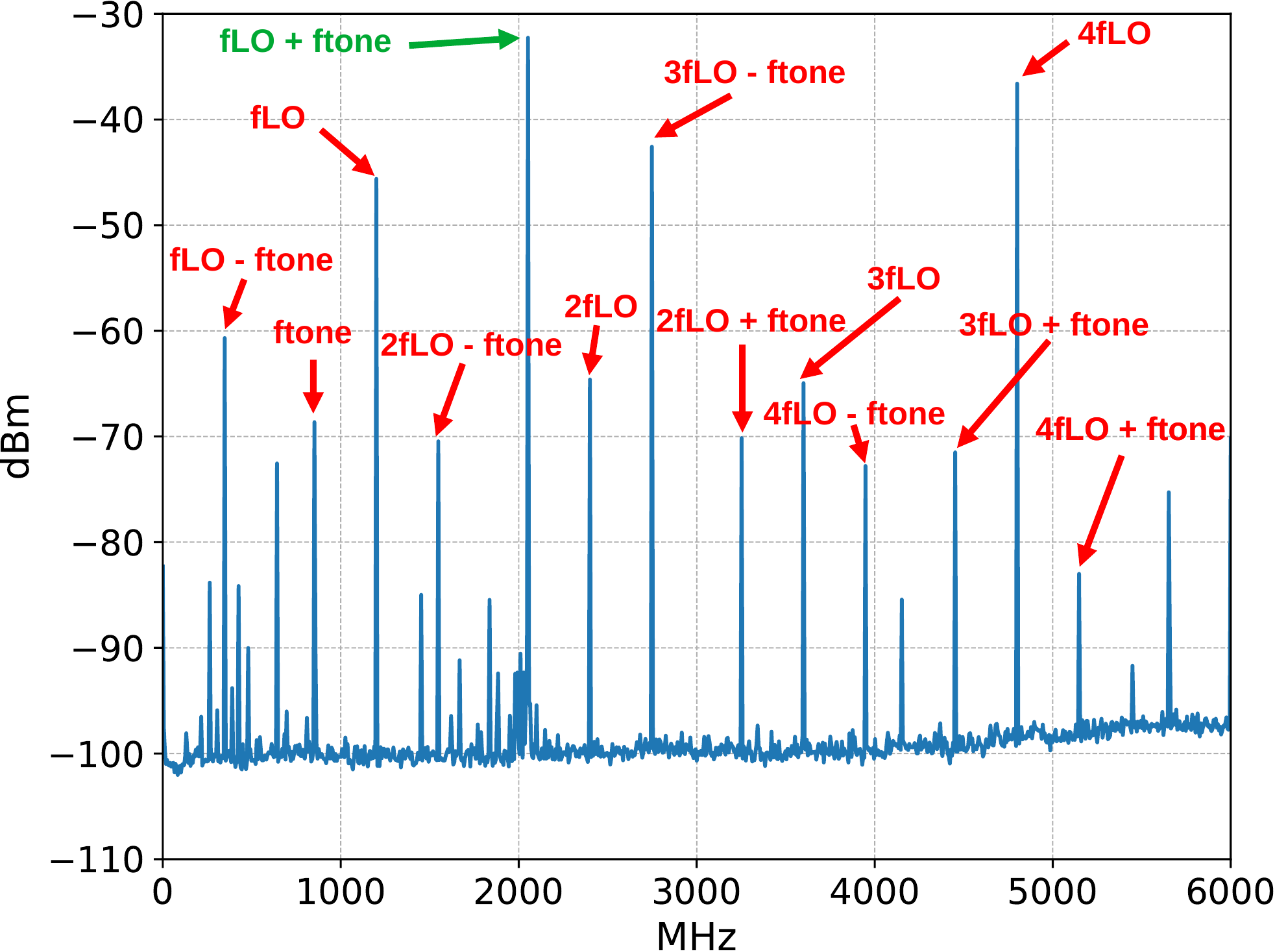}
\caption{Measured spectrum of ADL5375 Output.
    Input signals: \(f_{\text{tone}} = 850\,\text{MHz}\), \(f_{\text{LO}} = 1.2\,\text{GHz}\).}
    \label{fig:91MHz_quadra}
\end{figure}

Fig.~\ref{fig:91MHz_quadra} illustrates the measurement spectrum, captured by the spectrum analyzer (ZNL6), of a single tone signal scenario.
The modulated tone is visible at \( f_{\text{LO}} + f_{\text{tone}} \), corresponding to 2050\,MHz. 
In addition, feedthrough components at \( f_{\text{LO}} \) and \( f_{\text{tone}} \), as well as the lower sideband image at \( f_{\text{LO}} - f_{\text{tone}} \), are observed. 
Intermodulation products (IMDs) of the form \( n f_{\text{LO}} \pm f_{\text{tone}} \) are also present. 
Most of these spurs appear at power levels that, although lower than that of the excitation tone, remain non-negligible.
In the most extreme case, the component at \( 4 f_{\text{LO}} \) reaches a power level comparable to that of the main modulated tone at \( f_{\text{LO}} + f_{\text{tone}} \).

\subsubsection{Modeling}\label{subsec:IQ_model}

Due to the absence of detailed information regarding the internal architecture of the IQ modulator, developing a low-level architectural model is not feasible.
Instead, we adopt an empirical approach, constructing a behavioral model based on the fundamental principle of IQ modulation, as described by Equation~\ref{eq:first_equation}.

\begin{equation}
\begin{split}
IQ(t) &= S_I(t) \cdot LO_Q(t) + S_Q(t) \cdot LO_I(t) \\
\end{split}
\label{eq:first_equation}
\end{equation}

A simulation of Eq.~\ref{eq:first_equation} yields an ideal spectrum, where only the desired upper sideband appears at \( f_{\text{LO}} + f_{\text{tone}} \). 
However, experimental measurements reveal the presence of multiple spurious components.

We therefore progressively refine the model by introducing the dominant non-idealities observed experimentally. 
As a first step, to reproduce the lower sideband image at $f_{\text{LO}} - f_{\text{tone}}$, we introduce an amplitude imbalance between the I and Q components of the LO signal.
This imbalance reflects the behavior of the implemented modulator, which internally splits the LO signal into LO\textsubscript{I} and LO\textsubscript{Q}—a process that is inherently imperfect.
The imbalance is modeled by assigning distinct amplitudes to LO\textsubscript{I} and LO\textsubscript{Q}, denoted \( A_{\text{nLO}_I} \) and \( A_{\text{nLO}_Q} \), respectively, as shown in Equation~\ref{eq:expanded2_equation}.
When incorporated into the model, this results in a simulated spectrum that includes both the expected upper sideband (\( f_{\text{LO}} + f_{\text{tone}} \)) and the undesired lower sideband (\( f_{\text{LO}} - f_{\text{tone}} \)).

Furthermore, our measurements reveal IMDs at $a \cdot f_{\text{LO}} \pm f_{\text{tone}}$, where $a = 2$ to $4$.
To emulate these artifacts, we include in the model the imbalanced higher-order LO harmonics as shown in Equation~\ref{eq:expanded2_equation}. 

By using the mathematical identity \(\sin(a) \cdot \cos(b) = \frac{1}{2} \left( \sin(a+b) + \sin(a-b) \right)\), we expand the Equation~\ref{eq:expanded2_equation}.
At the end, we obtain all the frequency components at \( n \cdot f_{\text{LO}} \pm f_{\text{tone}} \), which include the non-rejected sideband image, the high sideband image, and the IMDs.

\begin{equation}
\begin{split}
IQ(t) &= A_{tone} \cos(w_{tone} \cdot t) \sum_{n=1}^{4} A_{nLO_Q} \sin(w_{nLO} \cdot t) + A_{tone} \sin(w_{tone} \cdot t) \sum_{n=1}^{4} A_{nLO_I} \cos(w_{nLO} \cdot t) \\
&= A_{tone} \cos(w_{tone}\cdot t) \cdot [A_{LO_Q} \sin(w_{LO} \cdot t)+A_{2LO_Q} \sin(w_{2LO} \cdot t)+A_{3LO_Q} \sin(w_{3LO} \cdot t)\\
&\quad +A_{4LO_Q} \sin(4w_{LO} \cdot t)] + A_{tone} \sin(w_{tone}\cdot t) \cdot [A_{LO_I} \cos(w_{LO}\cdot t)+A_{2LO_I} \cos(w_{2LO}\cdot t)\\
&\quad +A_{3LO_I} \cos(w_{3LO}\cdot t)+A_{4LO_I} \cos(w_{4LO}\cdot t)]\\
&= \frac{1}{2} [A_{tone} \cdot (A_{LO_Q}+A_{LO_I}) \sin( (w_{tone}+w_{LO})t) + A_{tone} \cdot (A_{LO_Q}+A_{LO_I}) \sin((w_{S}-w_{LO} )t )\\
&\quad + A_{tone} \cdot (A_{2LO_Q}+A_{2LO_I}) \sin((w_{tone}+w_{2LO})t ))  + A_{tone} \cdot (A_{2LO_Q}+A_{2LO_I}) \sin((w_{tone}-w_{2LO} )t) \\
&\quad + A_{tone} \cdot (A_{3LO_Q}+A_{3LO_I}) \sin( (w_{tone}+w_{LO})t) + A_{tone} \cdot (A_{3LO_Q}+A_{3LO_I}) \sin((w_{tone}-w_{LO} )t )\\
&\quad + A_{tone} \cdot (A_{4LO_Q}+A_{4LO_I}) \sin((w_{tone}+w_{2LO})t )+ A_{tone} \cdot (A_{4LO_Q}+A_{4LO_I}) \sin((w_{tone}-w_{2LO} )t )
]
\end{split}
\label{eq:expanded2_equation}
\end{equation}

The model parameters—specifically, the amplitudes of the imbalanced LO harmonics—are estimated from the measured spectrum shown in Fig.~\ref{fig:91MHz_quadra} by solving a system of equations detailed in~\cite{abdkrimi2025phd}. 
The resulting model takes as input the injected tone amplitude \(A_{\text{tone}}\) and angular frequency \(\omega_{\text{tone}}\).

However, as derived in Eq.~\ref{eq:expanded2_equation}, this analytical formulation does not reproduce all spectral components observed experimentally. 
First, fixed-frequency components at \(f_{\text{LO}}\), \(2f_{\text{LO}}\), \(3f_{\text{LO}}\), and \(4f_{\text{LO}}\) are visible in the measurements (see Fig.~\ref{fig:91MHz_quadra}). Their amplitudes are independent of both \(A_{\text{tone}}\) and \(\omega_{\text{tone}}\), and are therefore added to the \(IQ(t)\) signal resulting from Eq.~\ref{eq:expanded2_equation} as sinusoidal terms with amplitudes extracted from measurements.
Second, a baseband feedthrough component is consistently observed with an amplitude approximately equal to \(0.0112\,A_{\text{tone}}\). This feedthrough term is likewise incorporated into the \(IQ(t)\) signal within the behavioral model.
The complete model is described in Algorithm~\ref{alg:iq_modulation}.

\begin{algorithm}[h]
\caption{IQ Modulator Model}
\label{alg:iq_modulation}
\KwData{
$F_{\text{s}}$: Sampling frequency, 
$N$: Signal length, 
$A_{\text{tone}}$: Input tone amplitude, 
$A_{k\text{LO\_I}}$: LO in-phase harmonic amplitudes, 
$A_{k\text{LO\_Q}}$: LO quadrature harmonic amplitudes, 
$A_{k\text{LO}}$: Amplitudes of LO feedthrough harmonics, 
$f_{\text{tone}}$: Input tone frequency, 
$f_{\text{LO}}$: LO frequency
}

$w \leftarrow 2 \times \pi \times f_{\text{tone}}$ \;
$w_{\text{LO}} \leftarrow 2 \times \pi \times f_{\text{LO}}$ \;

$t \leftarrow$ array of time values from 0 to $N/F_{\text{s}}$ with spacing $1/F_{\text{s}}$\;

\For{$i$ from 1 to length of $t$}{
    $LO\_I \leftarrow \sum_{k=1}^{4} A_{k\text{LO\_I}} \times \cos(k \cdot w_{\text{LO}} \cdot t[i])$\;
    $LO\_Q \leftarrow \sum_{k=1}^{4} A_{k\text{LO\_Q}} \times \sin(k \cdot w_{\text{LO}} \cdot t[i])$\;
    $S\_I \leftarrow A_{\text{tone}} \times \cos(w \cdot t[i])$\;
    $S\_Q \leftarrow A_{\text{tone}} \times \sin(w \cdot t[i])$\;
    $S\_IQ \leftarrow S\_I \times LO\_Q + S\_Q \times LO\_I$\;
}

\For{$i$ from 1 to length of $t$}{
    $S\_IQ \leftarrow S\_IQ + 0.0112 \times A_{\text{tone}} \times \cos(w \cdot t[i]) + \sum_{k=1}^{4} A_{k\text{LO}} \times \sin(k \cdot w_{\text{LO}} \cdot t[i])$\;
}
\end{algorithm}

\subsubsection{Simulation and analysis}
\label{subsec:modul_comparanal}

To evaluate the accuracy and limitations of the proposed model, a comparative analysis is performed for two distinct scenarios: one with a high-frequency baseband tone (850\,MHz), which was used to find the model parameters, and another with a low-frequency baseband tone (90\,MHz).
In both cases, the tones are modulated by a 1200\,MHz LO, as is typically done in practice for reading out CONCERTO MKID arrays.

The developed model represents an analog component, and since the characterization was performed using the ZNL6, which samples at 12\,GHz and allows observation of spurs up to 6\,GHz, the model is also simulated with a sampling frequency of 12\,GHz.

\begin{figure}[h]
    \begin{minipage}{0.5\textwidth}
        \centering
        \includegraphics[width=\textwidth]{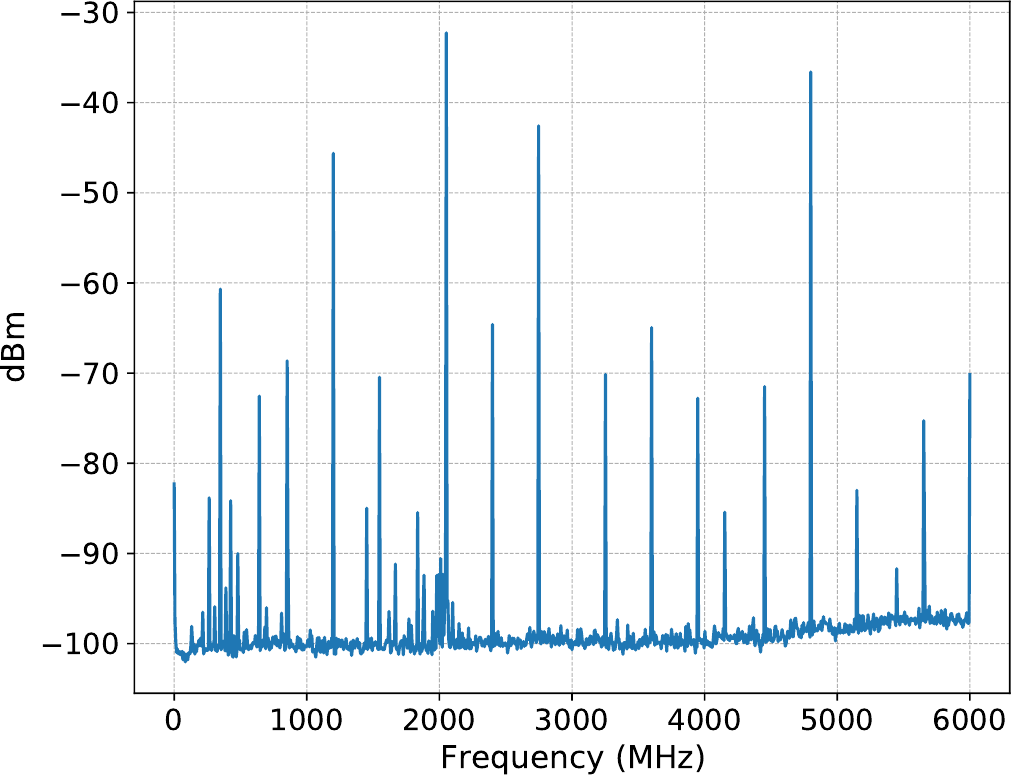}
    \end{minipage}%
    \begin{minipage}{0.5\textwidth}
        \centering
        \includegraphics[width=\textwidth]{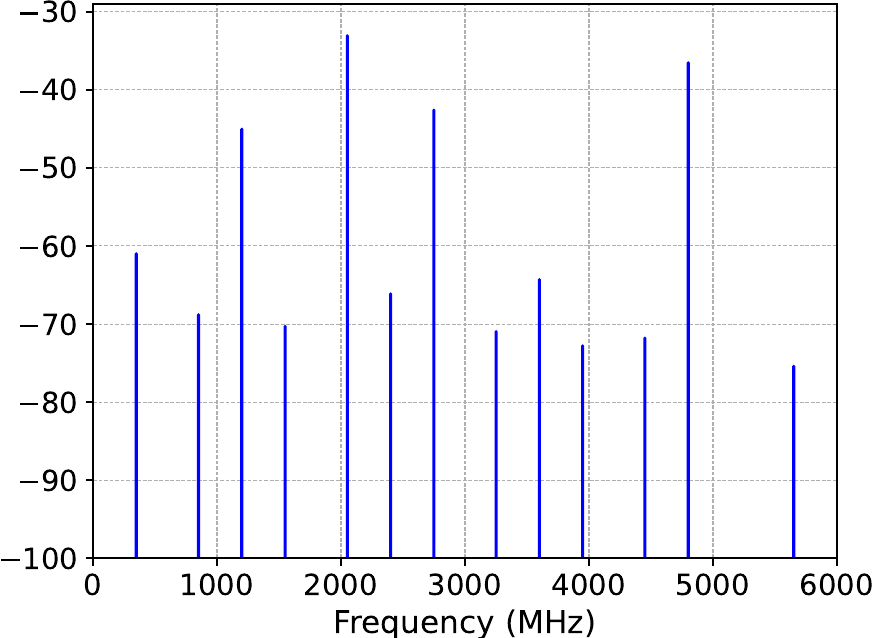}
    \end{minipage}
    \caption{Left: Measured spectrum at the modulator output. Baseband tone (\(f_{\text{tone}}\)) = 850\,MHz, LO sinusoid (\(f_{\text{LO}}\)) = 1.2\,GHz.
    Right: Simulation result of the modulator model with same input frequencies.}
    \label{fig:851MHz_combined}
\end{figure}

Fig.~\ref{fig:851MHz_combined} shows the output spectrum of the IQ modulator for an input tone at 850\,MHz.
The model and the measurement are in excellent agreement.  
All spectral components—including the modulated signal at 2050\,MHz (i.e., 850 + 1200)—match in both frequency and power.  
This match is expected, as this spectrum was used to tune the model parameters.

\begin{figure}[H]
    \begin{minipage}{0.5\textwidth}
        \centering
        \includegraphics[width=\textwidth]{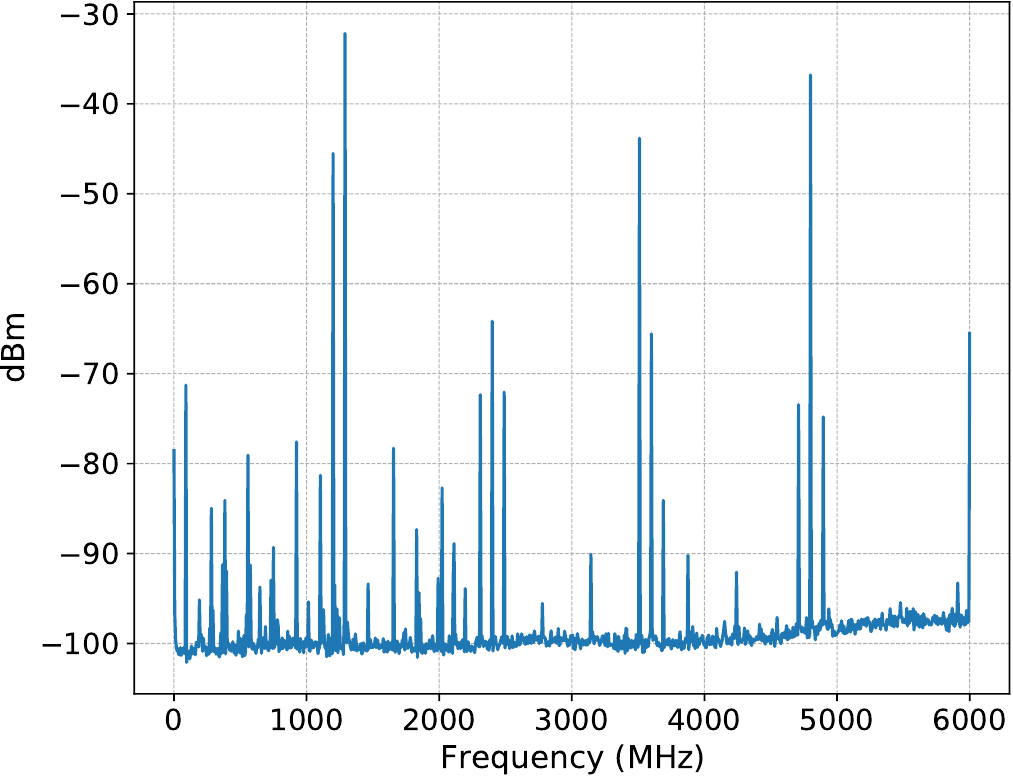}
    \end{minipage}%
    \begin{minipage}{0.54\textwidth}
        \centering
        \includegraphics[width=\textwidth]{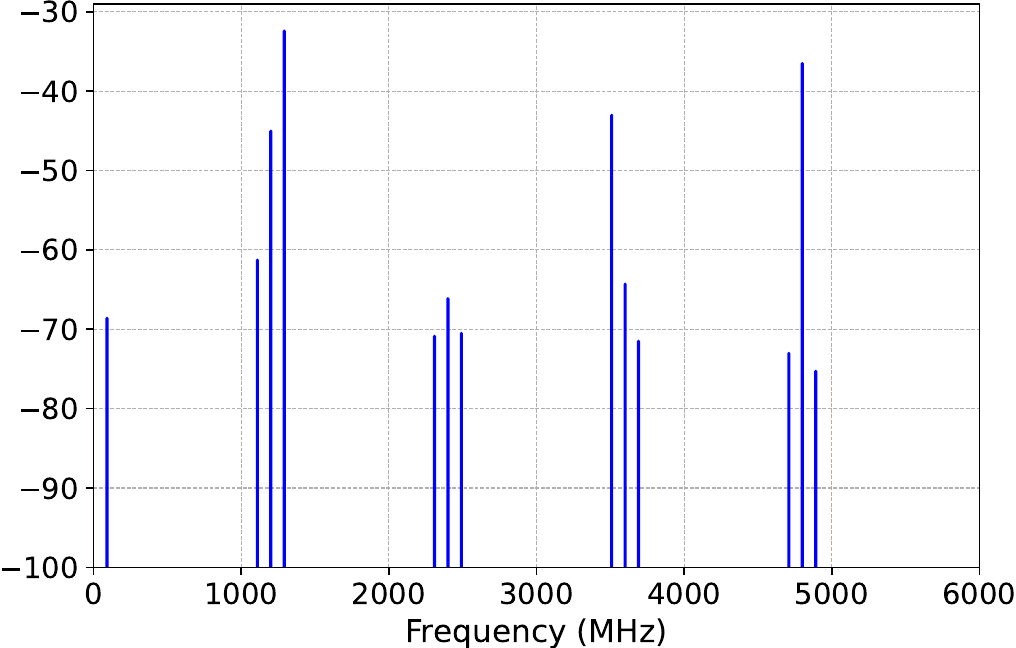}
    \end{minipage}
    \caption{Left: Measured spectrum at the modulator output. Baseband tone (\(f_{\text{tone}}\)) = 90\,MHz, LO sinusoid (\(f_{\text{LO}}\)) = 1.2\,GHz.
    Right: Simulation result of the modulator model with same input frequencies.}
    \label{fig:91MHz_combined}
\end{figure}

Fig.~\ref{fig:91MHz_combined} presents the measured and modeled spectra for a 90\,MHz input tone.  
The spur frequencies show excellent agreement between the model and the measurements.  
The spur power levels also generally match well, with two notable discrepancies observed for two specific IMDs.  
The first discrepancy occurs at \( f_{\text{LO}} - f_{\text{tone}} \), located at 1110\,MHz (1200\,MHz - 90\,MHz), where the model overestimates the power by approximately 20\,dB.  
The second is at \( 3f_{\text{LO}} + f_{\text{tone}} \), located at 3690\,MHz (3600\,MHz + 90\,MHz), where the model overestimates the power by about 12\,dB.

These differences are due to the fixed model parameters, which do not fully account for frequency-dependent behavior. 

The characterization and modeling work presented for the IQ modulator demonstrates the effectiveness of our methodology in capturing the essential non-linear characteristics of the ADL5375. 
By developing a behavioral model that incorporates LO imbalanced harmonics we have successfully reproduced the intermodulation distortion patterns observed in measurements, which degrade the MKIDs excitation signal quality. 
Moreover, improving the model by incorporating frequency-dependent behavior is a promising perspective for future work.
Finally, although the readout electronics must provide flexibility to adjust the LO frequency—allowing the frequency comb to be shifted slightly upward or downward depending on the specific MKID feedline—it remains possible to suppress undesired spectral components that lie far from the typical operating frequency band of the MKIDs. 
In particular, frequency components below 0.5\,GHz, as well as high-frequency components in the 3–6\,GHz range, can be filtered, thereby reducing the amount of unwanted noise injected into the MKID feedline.

\subsection{Mixer}
\subsubsection{Principle and Limitations}

The mixer demodulates the signal returning from the detector covering the frequency band of the MKIDs feedline \( [1.2-2.2 \, \text{GHz}] \) back to \( [0-1 \, \text{GHz}] \).
The mixer multiplies the incoming signal S with the LO. 
This multiplication creates two new frequency components: a high-band image and a low-band image. 
The high-band image frequency is the sum of LO and S frequencies, while the low-band image results from their difference.
Since the goal here is demodulation, the focus is on the low-band image, which contains the demodulated signal S.
To isolate the desired signal, a low-pass filter with a cutoff frequency of 1\,GHz is employed at the mixer's output to remove the unwanted high-band images.

Similar to the IQ modulator, the primary imperfection in mixers arises from the nonlinearities of the analog multiplier, which generate IMDs. 
However, since the mixer operates with single-ended signals, IQ imbalance is not a concern.
Furthermore, unlike the excitation chain, a low-pass filter with a 1\,GHz cutoff is positioned at the mixer output.

Nevertheless, if some of the IMD products fall within the [0--1\,GHz] baseband range, they will pass through this filter, reach the ADC, and propagate to the digital analysis stage, potentially corrupting the amplitude and phase extraction of neighboring tones.


\subsubsection{Simulation and Analysis}
\label{subsec:mixer_exper}

Similarly to the methodology adopted for the IQ modulator, the mixer is modeled starting from its fundamental mixing equation, which is then progressively enriched by incorporating the dominant non-idealities. The resulting behavioral model is validated through comparison with measurements.
A detailed description of this modeling procedure is provided in~\cite{abdkrimi2025phd}.

\begin{figure}[h]
    \centering
    \includegraphics[width=1\textwidth]{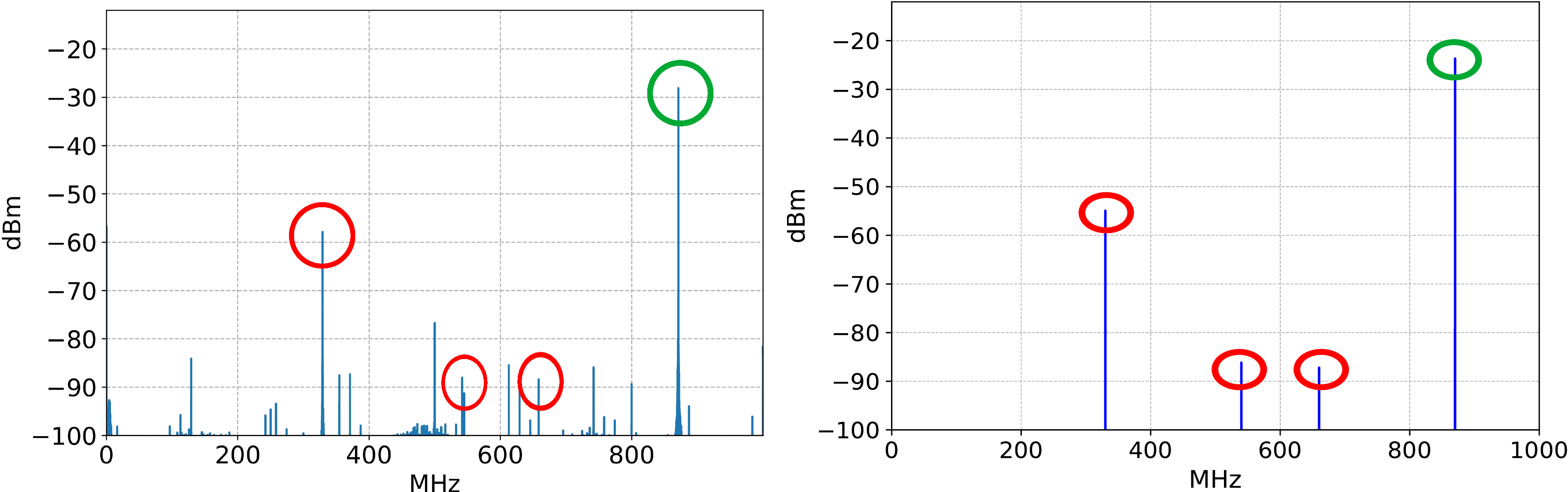}
    \caption{Left: Measured spectrum. Input : Modulated tone (\(f_{\text{tone}}+f_{\text{tone}}\)) = 870\,MHz+1200\,MHz, LO sinusoid (\(f_{\text{LO}}\)) = 1200\,MHz.
    Right: Simulation result of the modulator model with same input frequencies.  
    The green circle highlights the demodulated tone, while the red circles indicate the IMD components that match between the simulation and measurements.
 }
    \label{fig:comp_871}
\end{figure}

Here, we directly present a comparison between the simulated spectrum obtained from the model and the measured spectrum (see Fig.~\ref{fig:comp_871}).

As observed, three IMDs fall directly within the baseband frequency range [0–1]\,GHz. This is particularly critical because any IMD overlapping with the frequency of an analyzed tone will directly affect its extracted I/Q components. 

Thanks to the developed model, it becomes possible to predict which tones are impacted by IMDs. 
Since the frequencies of all injected tones are known a priori, the model allows us to determine the expected locations of their corresponding IMD products and identify which tones may experience degradation. 

\subsection{Analog-to-Digital Converter}

The ADC12D1000 is a 12-bit, 2\,GHz ADC.
It uses two interleaved ADCs, each sampling at 1\,GHz, with a 180-degree phase difference, effectively doubling the sampling frequency to 2\,GHz.
Like the DAC, the ADC shares the same imperfections, and supplementary ones due to the interleaved architecture.

\subsubsection{Digital crosstalk}\label{subsec:adc_crosstalk}

\begin{figure}[H]
    \centering
    \includegraphics[width=0.65\textwidth]{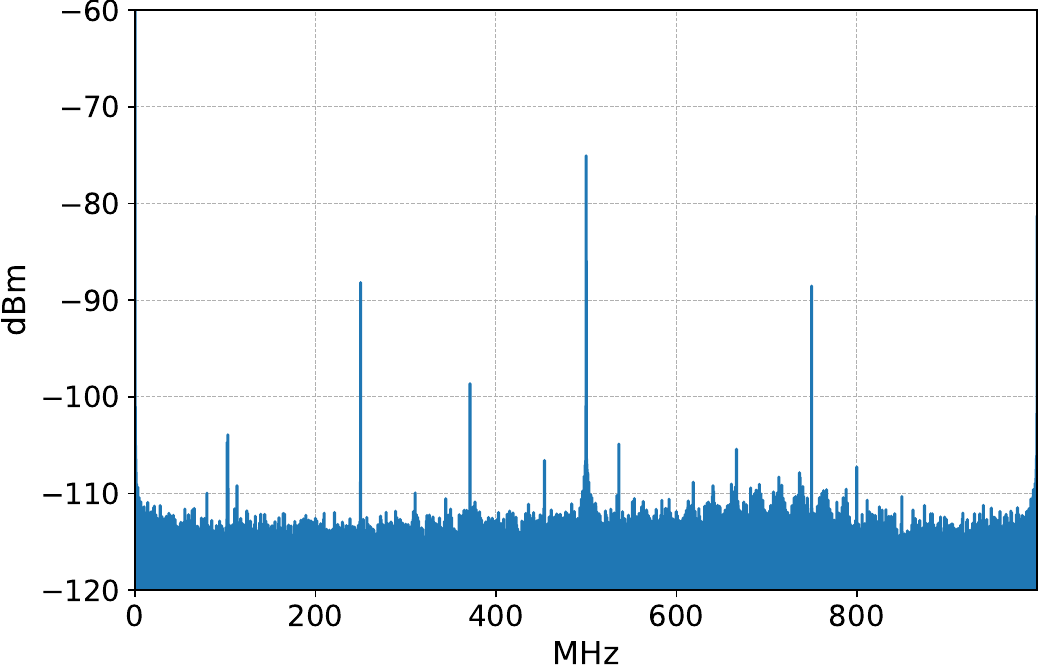}
    \caption{Spectrum of the ADC output with no input signal applied.
    The spectrum is obtained through FFT analysis of the digitized signal.
}
    \label{fig:0MHz_adc}
\end{figure}

With no input signal applied to the ADC, the output spectrum---shown in Fig.~\ref{fig:0MHz_adc}---reveals three prominent spurs at 250\,MHz, 500\,MHz, and 750\,MHz.
Since the ADC is not fed by a clock at these frequencies, these spurs likely originate from coupling at the board level, which includes a 250\,MHz clock for the FPGA and other digital components.
Although the exact origin of these spurs has not been identified, three sinusoids at the corresponding frequencies are simulated with amplitudes matching the measured power levels and incorporated into the ADC model.

\subsubsection{Interleaved architecture} \label{subsec:adc_spurs}

Interleaving ADCs enable higher sampling bandwidths by multiplying the sampling rate with the number of interleaved ADCs.
However, this architecture has imperfections, primarily due to mismatches in characteristics between ADCs, such as gain and offset voltage caused by fabrication imperfections, which introduce spurs in the ADC output spectrum.

\begin{figure}[h]
    \centering
    \includegraphics[width=0.85\textwidth]{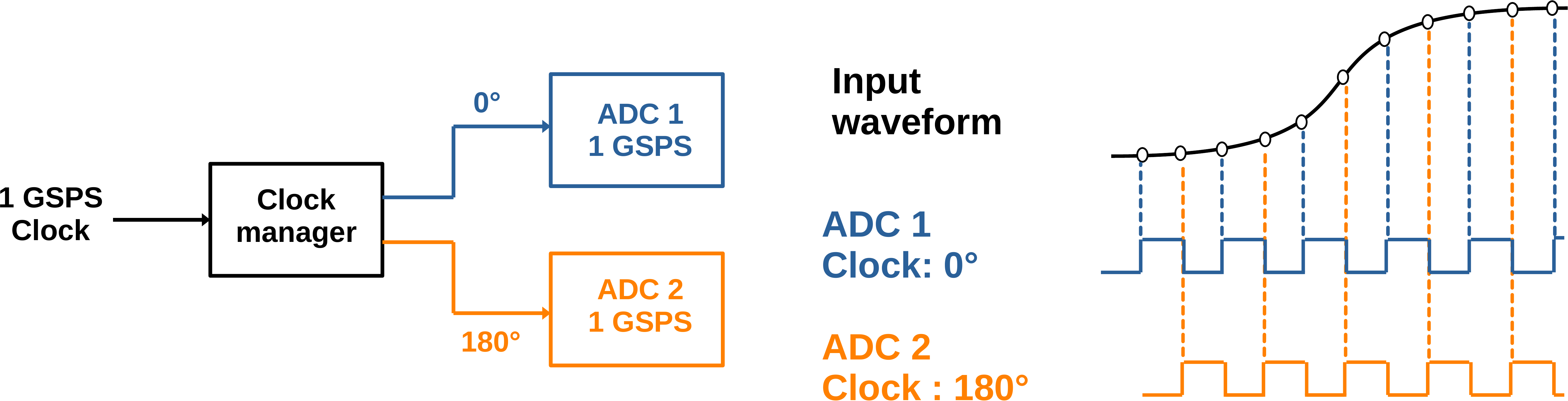}
    \caption{Illustration of the two ADCs interleaved architecture~\cite{harris2013abcs}.}
    \label{fig:inte}
\end{figure}

The ADC features two 1\,GHz interleaved ADCs and a clock manager. 
The clock manager is fed with a 1\,GHz clock and generates two 1\,GHz clocks phased 180 degrees apart, effectively doubling the overall sampling rate to \(F_s\)=2\,GHz (see Fig.~\ref{fig:inte}).

\begin{figure}[h]
    \centering
    \includegraphics[width=1\textwidth]{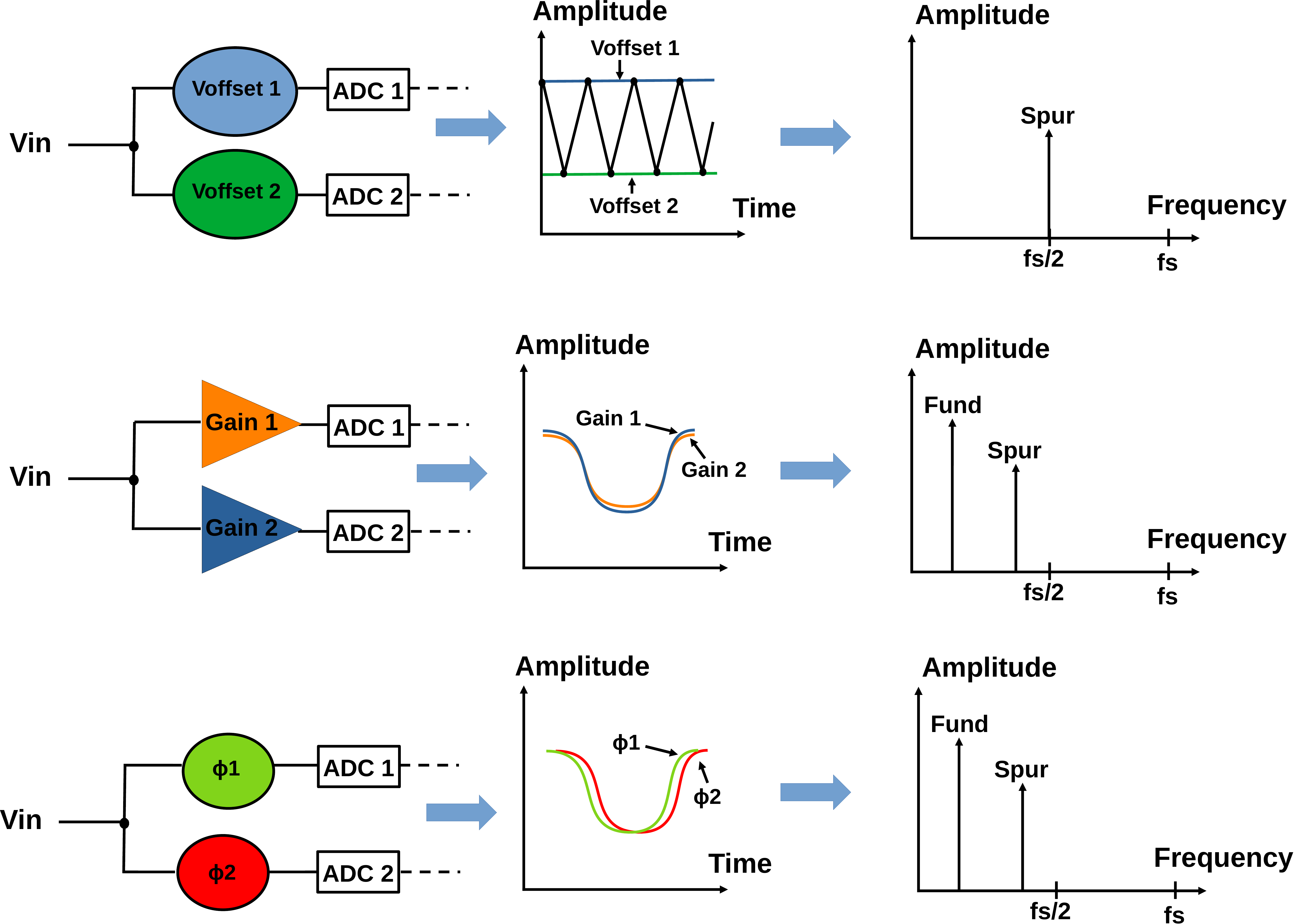}
    \caption{Illustration of the two ADCs interleaved architecture~\cite{harris2013abcs}.}
    \label{fig:spur_inter}
\end{figure}

The mismatches between the interleaved ADCs include voltage offset mismatch, gain mismatch, phase mismatch, and bandwidth mismatch~\cite{harris2013abcs}.
The voltage offset mismatch between the two ADCs causes sample values to alternate between the two ADCs' DC offsets, as illustrated in Fig.~\ref{fig:spur_inter}, resulting in a spur at \(F_s/2\), with the spur's magnitude depending on the mismatch magnitude, which is modeled by constants \(V_{\text{offset1}}\) and \(V_{\text{offset2}}\), appropriately added to the samples of ADC 1 and ADC 2, respectively.

Next, there is the gain mismatch, which introduces a spur at \(F_s/2 - f_{\text{tone}}\)~\cite{harris2013abcs}, with the spur power dependent on the mismatch amount, and it is modeled by gain components (\(V_{\text{Gain1}}\) and \(V_{\text{Gain2}}\)) which are multiplied by the samples of ADC 1 and ADC 2.

Regarding the phase mismatch, it represents a systematic phase offset between the clock rising edges of the two interleaved ADCs. 
Ideally, this offset would be 180°, but in practice, it tends to be 180° plus a constant \( \phi \).
This also introduces a spur at \(F_s/2 - f_{\text{tone}}\)~\cite{harris2013abcs}, whose power depends on both the phase mismatch value and the frequency of the tone fed to the ADC.
The phase offsets are added to the ideal time vector which is used to compute the tone, representing the process of sampling the signal at regular time intervals with a mismatched phase offset.

The ADC characterization setup allowed us to quantify of the model parameters discussed in the previous paragraph.
These components are adjusted to produce spur power levels consistent with the measurements. 
However, the last two mismatches induce spurs at the same frequency, \(F_s/2 - f_{\text{tone}}\).
To differentiate the contribution of each imperfection, near-DC tones are applied to the ADC for the gain mismatch. 
This approach is chosen because gain mismatch is unaffected by the tone's frequency, making it easier to distinguish from timing mismatch, which is evaluated using high-frequency tones.

\begin{figure}[h]
\centering
\begin{subfigure}[b]{0.49\textwidth}
    \centering
    \includegraphics[angle=0,width=\textwidth]{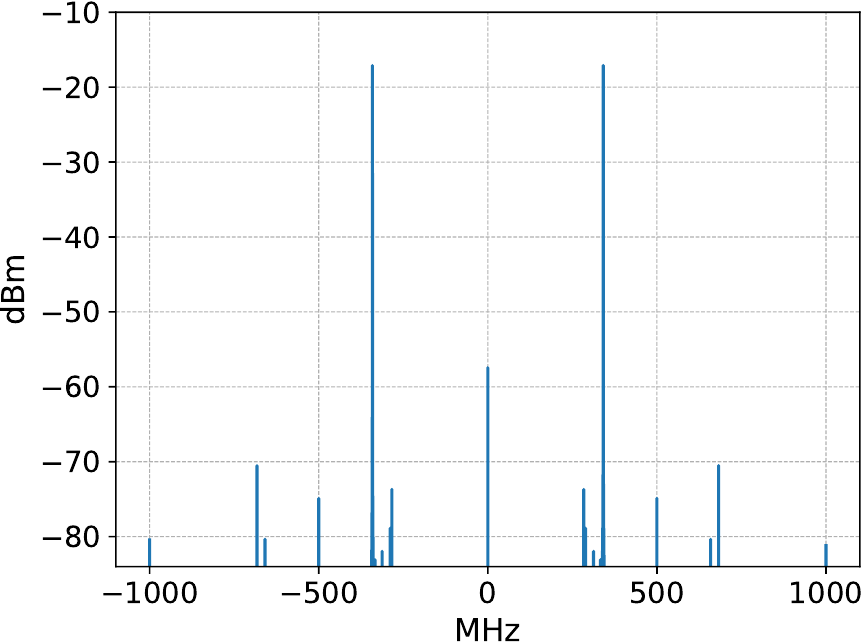}
    \caption{} 
\end{subfigure}
\hfill
\begin{subfigure}[b]{0.5\textwidth}
    \centering
    \includegraphics[angle=0,width=\textwidth]{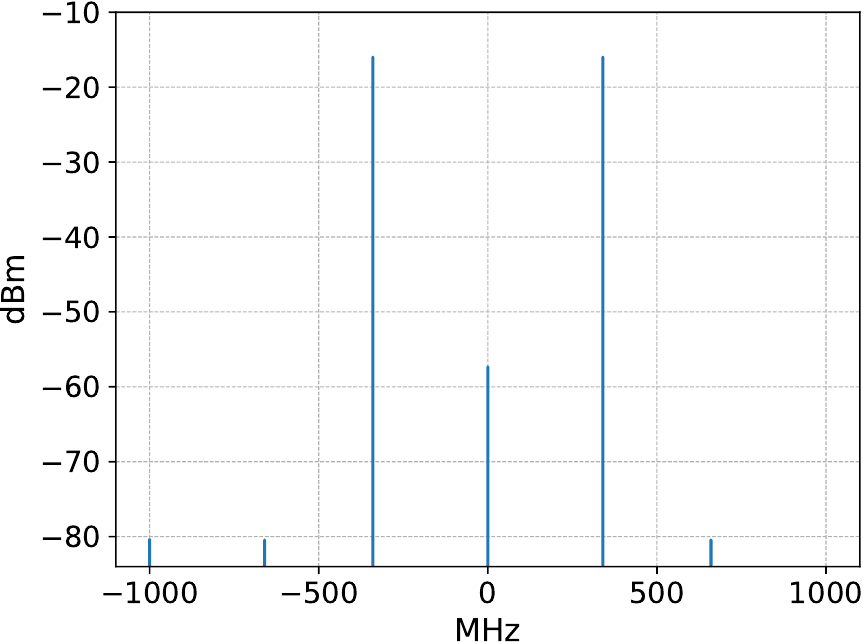}
    \caption{} 
\end{subfigure}
\caption{\label{fig:comp_340} (a) FFT of the ADC output, (b) FFT of the interleaved ADC model simulation. 
Input tone: 340\,MHz.
}
\end{figure}

Fig.~\ref{fig:comp_340} presents the spectrum for both the measurement and simulation.
The model parameters were adjusted to produce spurs that match the measurements and were found to be as follows: the offset voltage mismatch is 60\,$\mu$V, resulting in a spur level of -80\,dBm at \(F_s/2\). 
Additionally, the gain mismatch was tuned to 0.9995, and the timing mismatch to 0.5\,ps (the mismatch characteristics are not specified in the ADC datasheet), resulting in the \(F_s/2 - f_{\text{tone}}\) spur, occurring at 660\,MHz with a power level of -80\,dBm.
Lastly, the overall ADC offset voltage is measured at 400\,$\mu$V, corresponding approximately to 2~LSB, which is lower than the ADC datasheet specification of 5~LSB. 
This results in a spur at DC with a power of -57\,dBm.

Finally, the spur at 500\,MHz in the measurement spectrum in Fig.~\ref{fig:comp_340} is due to digital crosstalk, discussed in Section~\ref{subsec:adc_crosstalk}. 
The spur at 680\,MHz, with a power level of -70\,dBm, is the second harmonic of the tone and its origin is discussed in Section~\ref{subsec:adc_inl_dnl}.


\subsubsection{INL, DNL and white noise} \label{subsec:adc_inl_dnl}

\label{subsec:adc_inl_dnl}

In an ideal ADC, each digital output code corresponds to a uniform input voltage interval, with all intervals evenly distributed across the converter’s full-scale input range.
For a 12-bit ADC with a full-scale input of $0.8\,\text{V}_{\text{pp}}$, the ideal voltage step size is:
\begin{equation}
V_s
=
\frac{0.8~\mathrm{V}_{\mathrm{pp}}}{2^{12}}
\label{eq:voltage-step}
\end{equation}

Accordingly, each digital code maps to a unique voltage range of width $V_s$, and transitions between adjacent codes occur at threshold voltages given by:
\begin{equation}
V_{\mathrm{thresh,ideal}}[\mathrm{code}]
=
\mathrm{code}\cdot V_s
\label{eq:ideal-threshold-voltage}
\end{equation}

In practice, however, DNL causes deviations from these ideal threshold locations.
This effect is modeled by introducing a code-dependent error term $\text{DNL}_{\text{error}}[\text{code}]$, resulting in the actual threshold voltage:
\begin{equation}
V_{\mathrm{thresh,actual}}[\mathrm{code}]
=
\mathrm{code}\cdot V_s
+
\mathrm{DNL}_{\mathrm{error}}[\mathrm{code}]
\label{eq:actual-threshold-voltage}
\end{equation}

where $\text{DNL}_{\text{error}}$ is bounded within the range of $\left[-0.4 \cdot V_s,\; +0.4 \cdot V_s\right]$, such that the cumulative sum of DNL errors, which defines the INL, remains within the ADC12D1000's INL specification of $\pm 2.5$~LSB.

The developed model was simulated using a sinusoidal input signal with a representative frequency of 40\,MHz and a full-scale amplitude of $0.8\,\text{V}_{\text{pp}}$, consistent with the ADC’s dynamic range of $0.8\,\text{V}_{\text{pp}}$. 
The simulation was performed using a sampling frequency of 2000\,MHz—matching that of the ADC—and a frequency resolution of 8\,kHz.

To estimate the theoretical quantization noise floor, we use the formula derived in Section~\ref{subsec:dac_inl_dnl}:
\begin{equation}
\mathrm{Noise\ Floor}_{\mathrm{dBc}}
=
-6.02N - 1.76
-10\log_{10}\!\left(\frac{f_s}{2}\right)
+10\log_{10}(\Delta f)
\label{eq:quantization-noise-floor}
\end{equation}
where $N = 12$, $f_s = 2000$\,MHz, and $\Delta f = 8$\,kHz. Substituting these values yields:
\begin{equation}
\mathrm{Noise\ Floor}
=
-6.02\cdot 12
-1.76
-10\log_{10}(10^9)
+10\log_{10}(8000)
=
-125.05~\mathrm{dBc}
\label{eq:quantization-noise-floor-12bit}
\end{equation}

\begin{figure}[H]
    \centering
    \includegraphics[width=0.65\textwidth]{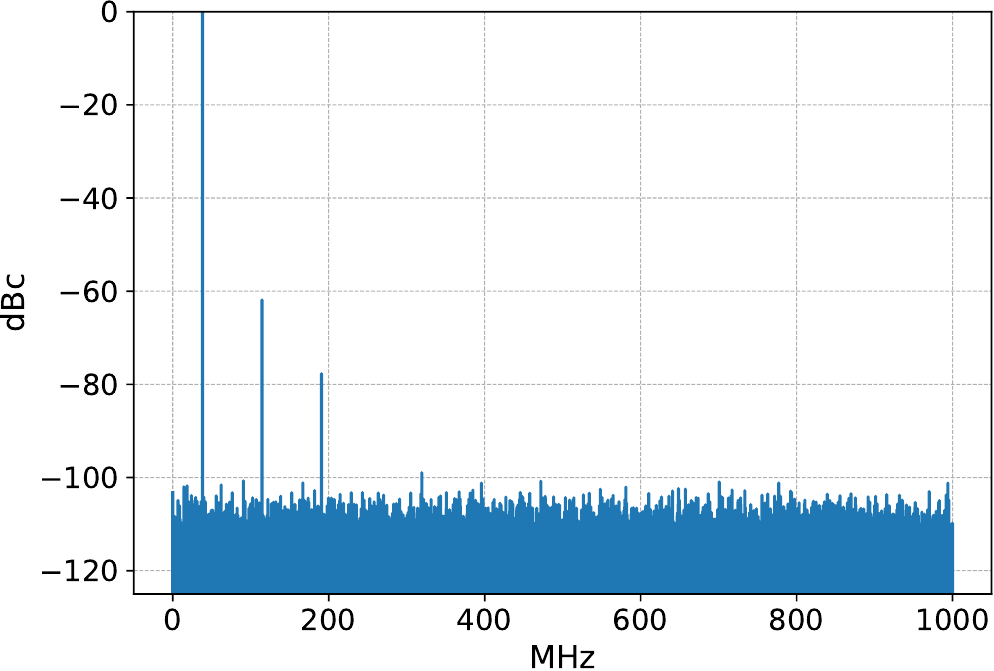}
    \caption{Output spectrum from the simulation of the developed ADC model incorporating DNL and INL imperfections.}
    \label{fig:ADC_INL}
\end{figure}

The model simulation results in a spectrum with an elevated noise floor, reaching approximately $-105\,\text{dBc}$, which is about 20\,dB above the theoretical quantization limit of $-125.05\,\text{dBc}$, as shown in Fig.~\ref{fig:ADC_INL}.

This elevated noise floor indicates that the ADC contributes more broadband noise than the DAC, whose noise floor, specifically at -124\,dBc is significantly lower (see Section~\ref{subsec:dac_inl_dnl}).
We further verified this experimentally by performing two loopback measurements: one through the full RF chain and one bypassing the RF stages. In both configurations, the measured noise floor remained unchanged, indicating that the ADC is the dominant white-noise source in the overall readout chain.

In summary, this characterization highlights two key points: (i) the interleaved ADC architecture introduces deterministic spurious components that pollute the spectrum, and (ii) the ADC noise floor dominates the broadband noise performance of the current readout system. 

At the time of component selection, high-speed converters combining multi-GS/s sampling rates with high resolution (particularly above 14 bits) were difficult to source. The available devices were typically implemented using time-interleaved architectures, meaning that multiple lower-speed ADC cores were combined to achieve a higher effective sampling rate. As a result, the observed mismatch-related effects are inherent to this class of converters and are difficult to avoid.

The proposed model therefore provides a practical short-term tool to predict the locations of additional spurious components and to assess whether they interfere with the I/Q components of an analyzed tone, in a manner similar to the approach used for the mixer.
In the longer term, this model enables the evaluation of alternative ADC choices—including different resolutions and interleaving architectures—in the context of next-generation readout electronics.

\section{Conclusion and perspectives}
\label{sec:conclusion}

In this paper, we presented a comprehensive characterization and behavioral modeling of the analog components of the KID\_READOUT electronics used in the CONCERTO instrument, namely the DAC, IQ modulator, mixer, and ADC.
By systematically analyzing each stage, we identified the dominant non-idealities and quantified their impact on the overall readout performance.

A first key outcome of this work is the determination of the maximum achievable multiplexing factor as a function of the number of bits allocated per tone, which is ultimately limited by DAC clipping rather than the DAC's slew-rate constraints. 
This result directly informs the trade-off between per-tone dynamic range and the number of simultaneously readable detectors, and therefore provides a concrete design guideline for future high-multiplexing MKID instruments.

Second, our noise analysis shows that the white noise floor of the current readout chain is dominated by the ADC, while the contributions from the DAC and RF stages remain sub-dominant. 
This clarifies that improving the converter choice or architecture is more impactful than refining the existing RF conditioning, and it offers a quantitative basis for evaluating alternative ADC technologies in next-generation designs.

Third, the developed behavioral models of the IQ modulator and mixer accurately reproduce the locations and amplitudes of intermodulation products and spurious tones observed in measurements. 
As all tone frequencies are known a priori, the model makes it possible to predict which readout channels are affected by specific IMD products, and thus to inform physicists of the vulnerable readout MKIDs in the frequency comb.

In future work, this analog simulator will be coupled to the existing digital twin of the FPGA-based DSP chain to form a complete end-to-end model of the MKIDs readout system, representing a first-of-its-kind framework.
This unified model will enable system-level studies and high-level design exploration of MKID readout architectures before hardware implementation, for example by assessing how changes in the DSP chain or replacing the ADC affect the performance of the entire electronics chain, from tone generation to the analyzed I/Q signals.


\bibliography{report_sample_bibtex}
\bibliographystyle{JHEP}

\end{document}